# Information Processing with Pure Spin Currents in Silicon: Spin Injection, Extraction, Manipulation and Detection

Olaf M. J. van 't Erve, Chaffra Awo-Affouda, Aubrey T. Hanbicki, Connie H. Li, Phillip E. Thompson, *member IEEE* and Berend T. Jonker

*Abstract*— We demonstrate that information can be transmitted and processed with pure spin currents in silicon. Fe/Al$_2$O$_3$ tunnel barrier contacts are used to produce significant electron spin polarization in the silicon, generating a spin current which flows outside of the charge current path. The spin orientation of this pure spin current is controlled in one of three ways: *(a)* by switching the magnetization of the Fe contact, *(b)* by changing the polarity of the bias on the Fe/Al$_2$O$_3$ "injector" contact, which enables the generation of either majority or minority spin populations in the Si, providing a way to electrically manipulate the injected spin orientation without changing the magnetization of the contact itself, and *(c)* by inducing spin precession through application of a small perpendicular magnetic field. Spin polarization by electrical extraction is as effective as that achieved by the more common electrical spin injection. The output characteristics of a planar silicon three terminal device are very similar to those of non-volatile giant magnetoresistance metal spin-valve structures.

## I. INTRODUCTION

THE *International Technology Roadmap for Semiconductors* has identified the electron's spin angular momentum as a new state variable that should be explored as an alternative to the electron's charge for use beyond CMOS [1]. The use of **pure spin currents** to process information is attractive because it potentially circumvents the constraints of capacitive time constants, resistive power dissipation and heat buildup which accompany charge motion. The generation of significant spin polarization is a basic requirement for utilizing spin as an alternate state variable in semiconductors. [2] This has typically been accomplished by electrically injecting spin-polarized electrons or holes from a magnetic contact with an intrinsic spin-polarized density of states and carrier population into the semiconductor host of choice. This has been demonstrated in GaAs using a variety of contact materials, including both magnetic semiconductors (*p*-GaMnAs, [3] *n*-ZnMnSe, [4],[5] *n*-CdCr$_2$Se$_4$ [6]) and ferromagnetic metals (Fe, [7],[8] FeCo, [9] FeGa, [10] MnAs [11]), and more recently with magnetic metals on Si. [12],[13],[14] Model calculations indicate that selectively *extracting* carriers of one spin state from the unpolarized carrier population of the semiconductor using the spin filtering provided by the spin-polarized density of states of the magnetic contact should be an equally effective avenue towards achieving high spin polarization in the semiconductor. [15],[16],[17] Evidence for this mechanism was presented for metals, [18],[19],[20] and for GaAs [21],[22] and graphene [23]. However, the contact bias dependence of the semiconductor spin polarization is not well understood, and remains a topic of considerable discussion. [22],[24],[25]

We have previously shown that an Fe/Al$_2$O$_3$ and an Fe Schottky tunnel barrier contact on Si can be used to electrically inject spin-polarized electrons from the Fe, resulting in large electron majority spin polarizations in the Si (referenced to the Fe magnetization).[12],[26] We describe here electrical spin *extraction* from Si using the same contact, where a positive voltage applied to the Fe results in an electron current from the Si into the Fe. The spin-polarized density of states at the Fe Fermi energy provides higher conductance for majority spin electrons, resulting in an accumulation of minority spin electrons in the Si, i.e. the net electron spin orientation in the Si is opposite that of the Fe "injector". The minority spin polarization in Si results in a splitting of the spin-dependent chemical potential, which is detected as a voltage at a second Fe/Al$_2$O$_3$ contact configured as a nonlocal detector.[27] These results show that either majority or minority spin populations can be generated in the Si simply by changing the polarity of the bias on the Fe/Al$_2$O$_3$ "injector" contact, providing a way to electrically change the injected spin orientation without changing the magnetization of the contact itself. We demonstrate nonlocal spin valve behavior as the magnetization of the injector and detector are switched from parallel to anti-parallel, and precession of the minority spin current generated by spin extraction. The contact bias and magnetization, together with precession, allow full control over the orientation of the spin in the silicon channel and subsequent detection as a voltage, demonstrating that information can be transmitted and processed with pure spin currents in silicon.

Manuscript received April 17, 2009. This work was supported by the office of Naval Research and core programs at the Naval Research Laboratory. O. M. J. van 't Erve gratefully acknowledges support as an NRL/George Washington University Research Associate, and Chaffra Awo-Affouda gratefully acknowledges support as an NRL/National Research Council Postdoctoral Associate.

The authors are with the Naval Research laboratory, 4555 Overlook Ave SW, 20375 Washington DC, USA

## II. FABRICATION

The samples consist of a 250 nm thick Si(001) epilayer grown by molecular beam epitaxy (MBE) at 1 Å/sec from an electron beam source at a substrate temperature of 500 °C on an undoped Si(001) substrate. The film was *n*-doped (phosphorus) at ~ $3 \times 10^{18}$ cm$^{-3}$ (25 °C), near the metal-insulator transition to avoid carrier freeze-out at lower temperatures. After a 10% HF acid etch and deionized water rinse to produce a hydrogen terminated surface, the samples were loaded into a second MBE system, heated to desorb the hydrogen, and a 1 nm $Al_2O_3$ tunnel barrier and 10 nm polycrystalline Fe film were deposited as described previously. [14]

Conventional photolithography and wet chemical etching were used to define multiple nonlocal spin valve (NLSV) structures, each consisting of four Fe contacts spaced as shown in Fig. 1. Contacts 2 (24 x 100 μm) and 3 (6 x 100 μm) are separated by 1 μm and serve as the spin detector and injector, respectively. The different aspect ratios provide different coercive fields with the easy axis along the long axis so that the contact magnetizations can be aligned parallel or anti-parallel. Contacts 1 and 4 are reference contacts (100 x 150 μm) located several spin diffusion lengths away from the transport channel to insure that the spin polarization is zero at these spin grounds.

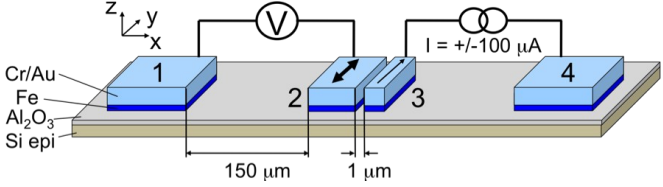

Fig. 1. Schematic layout of four terminal nonlocal device, a current is applied to contact 3 and 4, and a voltage is measured across contact 1 and 2.

## III. RESULTS

Fig. 2 shows the I-V curve at 15 K for these Fe/$Al_2O_3$/Si contacts. The modest temperature dependence of the normalized zero bias resistance, [28] defined as R(T)/R(300 K) (upper inset Fig. 2) and the good fits to the Brinkman, Dynes and Rowell model for asymmetric tunnel barriers [29] (lower inset Fig. 2) confirm the tunneling nature of these contacts. Tunnel contacts have been shown to provide the spin dependent interface resistance needed to create a spin polarization in a semiconductor [30],[31].

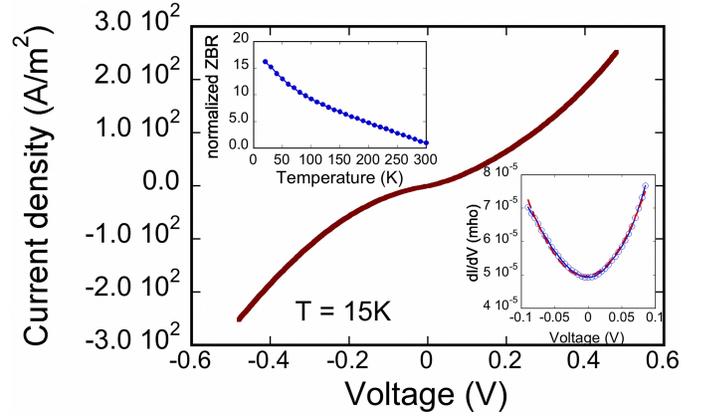

Fig. 2. I-V curve of the Fe/$Al_2O_3$/Si contacts, measured from Fe to a large Ohmic surface contact. The upper inset shows a modest change in zero bias resistance versus temperature. The lower inset shows the good fit of the data to the Brinkman, Dynes and Rowell model for asymmetric tunnel barriers.

In the spin *injection* experiment, a negative bias is applied to contact 3, and a spin-polarized electron current is injected from contact 3 into the Si to produce *(a)* a spin-polarized charge current, which flows to contact 4 due to the applied bias, and *(b)* a pure spin current, which diffuses isotropically. The electron spin is majority spin [12] oriented in the plane of the surface due to the in-plane magnetization of the Fe contacts (and opposite the magnetization of the injector by convention). We analyze the pure spin diffusion current at contact 2, which is outside of the charge current path ("nonlocal"), where the spin polarization in the Si results in a splitting of the electrochemical potential (Fig. 3). This is manifested as a voltage proportional to the projection of the semiconductor spin accumulation onto the magnetization direction of the detector contact.

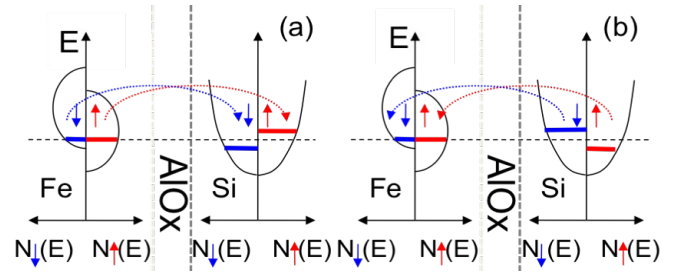

Fig. 3. Schematic illustration of (a) injection and (b) extraction of spins from Silicon by means of spin dependent tunneling.

Fig. 4 shows the nonlocal voltage as a function of an in-plane magnetic field $B_y$ that is used to switch the magnetization of the injector and detector contacts. At large negative field, the magnetizations are parallel, and the majority spins injected into the Si are parallel to the electron spin orientation of the detector, resulting in a minimum in the nonlocal voltage at contact 2.

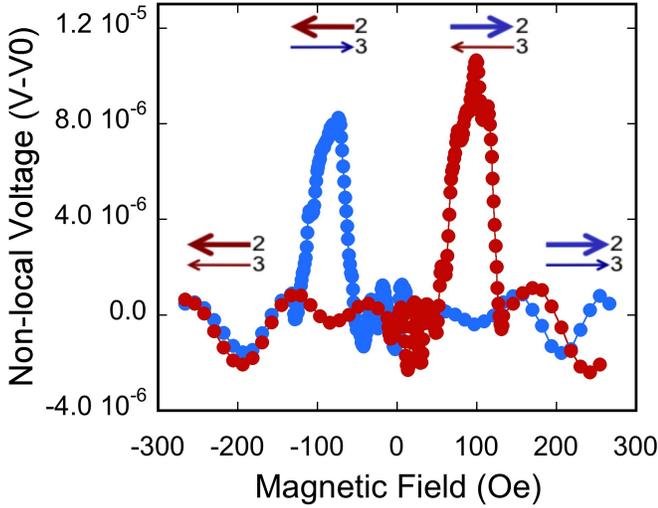

Fig. 4. Nonlocal voltage versus inplane magnetic field, for an injection current of -100 µA at 10 K. Two levels corresponding to the parallel and anti-parallel remanent states are clearly visible.

As the field is increased (red symbols), contact 2 with the lower coercive field switches so that the injector/detector magnetizations are anti-parallel. The orientation of the spin current diffusing to contact 2 is now anti-parallel to that of the detector, producing an abrupt increase at $B_y \sim 50$ Oe and a plateau in the nonlocal voltage. As $B_y$ is increased further, contact 3 also switches so that injector and detector contact magnetizations are again parallel, and the nonlocal voltage abruptly decreases to its minimum value. The process is repeated as the field is swept in the opposite direction (blue symbols), producing an output characteristic similar to that seen in metal pseudo spin-valve structures [32], and confirming remanent (non-volatile) parallel and anti-parallel contact orientations.

The top half of Fig. 5 shows similar data for decreasing values of the bias current flowing between contacts 3 and 4 (from -200 µA to -25 µA). As the bias voltage at the injector changes from negative to positive, electrons flow from the Si into the Fe contact (+50 µA to +150 µA). The higher conductivity of the Fe majority spin channel results in efficient majority spin extraction and accumulation of minority spin electrons in the Si, as illustrated in Fig. 3(b). A minority rather than majority pure spin current now diffuses from the injector to the detector contact 2, resulting in an inversion of the nonlocal voltage peaks for anti-parallel alignment of injector and detector contact magnetizations. The magnitude of the nonlocal voltage is roughly linear with the magnitude of the bias current, similar to references 20 to 23 for graphene and metals, but unlike [22] for GaAs where the non-monotonic behavior was attributed to localized electrons in bands near the surface due to the doping profile [24] or resonant states at the Fe/GaAs interface. [25]

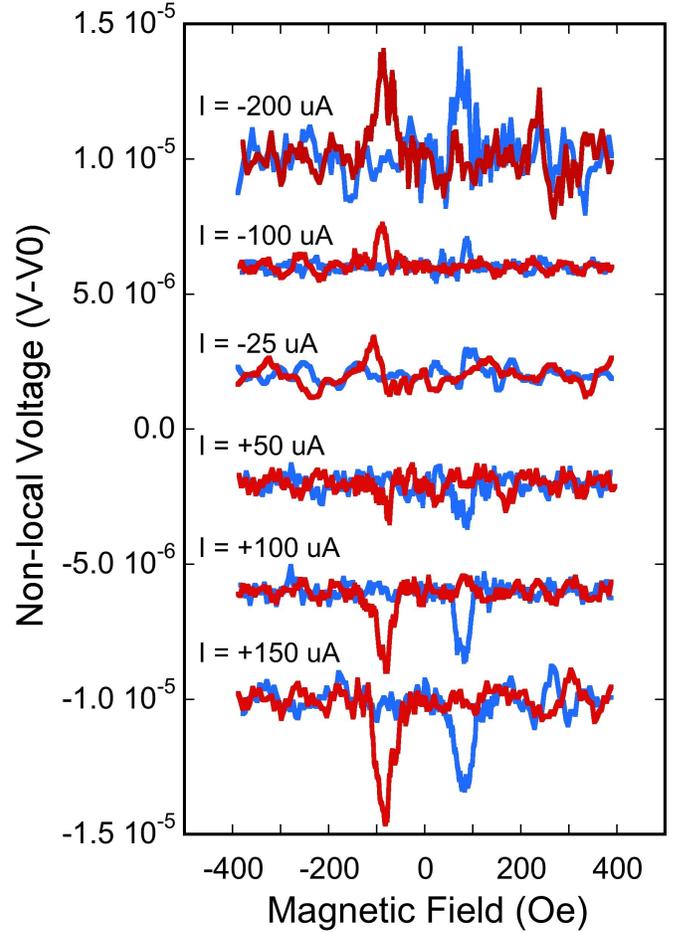

Fig. 5. Nonlocal voltage versus inplane magnetic field at 10 K for several values of the injection current, graphs are offset for clarity. For negative bias, electrons are injected from Fe into the Si channel and the change in nonlocal voltage is consistent with majority spin injection. For positive bias, electrons are extracted from the Silicon into the Fe contact. The majority spins are more readily extracted resulting in the accumulation of minority spin in the Silicon. A change in sign is seen for nonlocal voltage peaks for the antiparallel state, consistent with minority spin accumulation.

Hanle effect curves for bias currents of ±100 µA are shown in figure 6, demonstrating spin precession for both spin injection and extraction. The magnetizations of the injector and detector contacts are placed in the parallel remanent state, and a magnetic field (Hanle field, $B_z$) is applied perpendicular to the surface of the device. The lower trace is the Hanle curve for spin injection at a bias current of -100 µA. As the Hanle field increases from zero, the injected spins in the silicon precess around $B_z$ during transit to the detector contact, resulting in an increasing degree of anti-parallel alignment relative to the detector spin orientation and a corresponding increase in the nonlocal signal. This is consistent with the accumulation of majority spin in the Si channel. The upper trace is the Hanle curve for spin extraction at a bias of +100µA. Because minority spin is generated in the Si, the nonlocal voltage exhibits a maximum at $B_z = 0$ due to the anti-parallel alignment of the spin in the Si relative to the detector, and decreases with increasing Hanle field as spin precession leads to parallel alignment. Variations in transit times in the transport channel due to the width of the injector and detector contacts truncate the full Hanle curve and suppress further precessional oscillations in the nonlocal voltage.

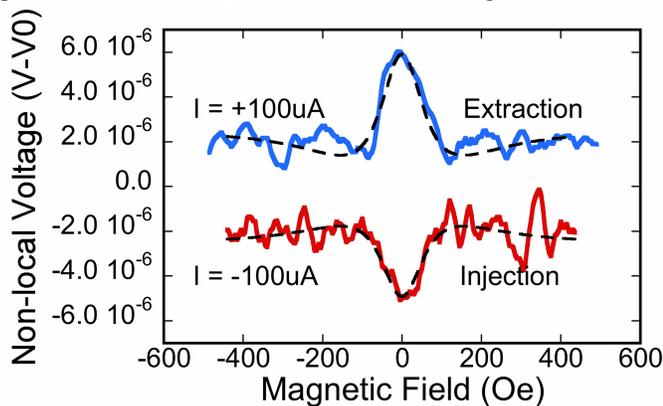

Fig. 6. Hanle measurements at 10 K for positive and negative injector current, with the graphs offset for clarity. The Hanle curve shows a dip or a peak for injection and extraction respectively, consistent with majority and minority spin accumulation. The dashed lines are fits to the Hanle data using g = 2, spin diffusion constant $D_s$ = 10 cm$^2$/s and spin lifetime $\tau_s$ = 0.9 ns.

The overall shape of the Hanle curve provides a measure of the spin lifetime in the Si channel. Fits to the data of Figure 6 using an approach similar to that of [22] yield spin lifetimes of ~ 1 ns. The lifetimes for majority and minority spins (referenced to the injector contact 3) in the Si channel are comparable, as expected for a non-magnetic semiconductor. We attribute the relatively short spin lifetimes to the fact that this lateral transport geometry probes spin diffusion near the Si/Al$_2$O$_3$ interface, where surface scattering and interface states are likely to produce more rapid spin relaxation than in the bulk. In addition, our Si epilayers are grown at significantly lower temperatures (500C) than those used in bulk growth, which results in higher defect densities.

## IV. Conclusion

In summary, we have demonstrated injection and extraction of spins from a Si channel by controlling the bias on the ferromagnetic injector contact, resulting in the accumulation of majority or minority spins in Si. This provides a simple electrical means to manipulate the spin orientation in future silicon-based spintronic devices without switching the magnetization of the contacts. An alternative method to electrically manipulate the spin orientation using two noncollinear injectors in the injection regime has been shown in metals [33]. The contact bias and magnetization, together with spin precession in the transport channel, allow full control over the orientation of the spin in the silicon channel and subsequent detection as a voltage. The ferromagnetic contacts provide non-volatile functionality as well as the potential for reprogrammability. This was accomplished in a lateral transport geometry using lithographic techniques compatible with existing device geometries and fabrication methods. Although the measurements reported here were performed at low temperature to reduce the noise arising from the relatively high Fe/Al$_2$O$_3$ contact resistance, techniques to reduce the interface resistance [34] or take advantage of fundamental band structure symmetries between Fe and Si [35] offer promising avenues to significantly improve the signal to noise for room temperature operation.

References

[1] International Technology Roadmap for Semiconductors, 2007 Edition. See http://www.itrs.net/Links/2007ITRS/Home2007.htm
[2] B. T. Jonker, "Progress toward electrical injection of spin-polarized electrons into semiconductors," *Proc. IEEE*, vol. 91, pp. 727-740, 2003.
[3] Y. Ohno, D. K. Young, B. Beschoten, F. Matsukura, H. Ohno and D.D. Awschalom, "Electrical spin injection in a ferromagnetic semiconductor heterostructure," *Nature,* vol. 402, pp. 790-792, 1999.
[4] R. Fiederling, M. Keim, G. Reuscher, W. Ossau, G. Schmidt, A. Waag and L. W. Molenkamp, "Injection and detection of a spin-polarized current in a light-emitting diode," *Nature*, vol. 402, pp. 787-790, 1999.
[5] B. T. Jonker, Y. D. Park, B. R. Bennett, H. D. Cheong, G. Kioseoglou, and A. Petrou, "Robust electrical spin injection into a semiconductor heterostructure," *Phys. Rev. B*, vol. 62, pp. 8180-8183, 2000.
[6] Y. D. Park, A. T. Hanbicki, J. E. Mattson, and B. T. Jonker, "Epitaxial growth of an n-type ferromagnetic semiconductor CdCr$_2$Se$_4$ on GaAs(001) and GaP(001),", *Appl. Phys. Lett.,* vol. 81, pp. 1471-1473, 2002.
[7] A. T. Hanbicki, B. T. Jonker, G. Itskos, G. Kioseoglou and A. Petrou, "Efficient electrical spin injection from a magnetic metal/tunnel barrier contact into a semiconductor," *Appl. Phys. Lett.,* vol. 80, pp. 1240-1242, 2002.
[8] V. F. Motsnyi, J. De Boeck, J. Das, W. Van Roy, G. Borghs, E. Goovaerts, and V. I. Safarov, "Electrical spin injection in a ferromagnet/tunnel barrier/semiconductor heterostructure," *Appl. Phys. Lett.,* vol. 81, pp. 265-267, 2002.
[9] X. Jiang, R. Wang, R. M. Shelby, R. M. Macfarlane, S. R. Bank, J. S. Harris and S. S. P. Parkin, "Highly Spin-Polarized Room-Temperature Tunnel Injector for Semiconductor Spintronics using MgO(100)," *Phys. Rev. Lett.,* vol. 94, p. 056601, 2005.
[10] O. M. J. van 't Erve, C. H. Li, G. Kioseoglou, A. T. Hanbicki, M. Osofsky, S.-F. Cheng, and B. T. Jonker, "Epitaxial growth and electrical spin injection from Fe$_{(1-x)}$Ga$_x$ (001) films on AlGaAs/GaAs (001) heterostructures," *Appl. Phys. Lett.,* vol. 91, p. 122515, 2007.
[11] M. Ramsteiner, H. Y. Hao, A. Kawaharazuka, H. J. Zhu, M. Kastner, R. Hey, L. Daweritz, H. T. Grahn and K. H. Ploog, " Electrical spin injection from ferromagnetic MnAs metal layers into GaAs," *Phys. Rev. B,* vol. 66, p. 081304(R), 2002.
[12] B. T. Jonker, G. Kioseoglou, A. T. Hanbicki, C. H. Li and P. E. Thompson, "Electrical spin-injection into silicon from a ferromagnetic metal/tunnel barrier contac," *Nature Phys.,* Vol. 3, pp. 542-546, 2007
[13] I. Appelbaum, B. Huang and D. Monsma, "Electronic measurement and control of spin transport in silicon," *Nature*, vol. 447, pp. 295-298, 2007.
[14] O. M. J. van 't Erve, A.T. Hanbicki, M. Holub, C.H. Li, C. Awo-Affouda, P.E. Thompson and B.T. Jonker, ""Electrical injection and detection of spin-polarized carriers in silicon in a lateral transport geometry," *Appl. Phys. Lett.*, vol. 91, p. 212109, 2007.
[15] I. Zutic, J. Fabian and S. Das Sarma, "Spin-Polarized Transport in Inhomogeneous Magnetic Semiconductors: Theory of Magnetic/Nonmagnetic p-n Junctions," *Phys. Rev. Lett.,* vol. 88, p. 066603, 2002.
[16] A. M. Bratkovsky and V. V. Osipov, "Efficient spin extraction from nonmagnetic semiconductors near forward-biased ferromagnetic-semiconductor modified junctions at low spin polarization of current," *J. Appl. Phys.*, vol. 96, pp. 4525-4529, 2004.
[17] V. V. Osipov, A. G. Petukhov and V. N. Smelyanskiy, "Complete spin polarization of electrons in semiconductor layers and quantum dots," *Appl. Phys. Lett.*, vol. 87, p. 202112, 2005.
[18] M. Johnson and R. H. Silsbee, "Interfacial charge-spin coupling: Injection and detection of spin magnetization in metals," *Phys. Rev. Lett.,* vol. 55, pp. 1790-1793, 1985.
[19] F. J. Jedema, A. T. Filip, and B. J. van Wees, "Electrical spin injection and accumulation at room temperature in an all-metal mesoscopic spin valve," *Nature,* vol. 410, pp. 345-348, 2001.
[20] S. O Valenzuela, D. J. Monsma, C. M. Marcus, V. Narayanamurti and M. Tinkham, "Spin Polarized Tunneling at Finite Bias," *Phys. Rev. Lett.,* vol. 94, p. 196601, 2005.
[21] S. A. Crooker, M. Furis, X. Lou, C. Adelmann, D. L. Smith, C. J. Palmstrøm, and P. A. Crowell, "Imaging Spin Transport in Lateral Ferromagnet/Semiconductor Structures," *Science,* vol. 309, pp. 2191-2195, 2005.

[22] X. Lou, C. Adelmann, S. A. Crooker, E. S. Garlid, J. Zhang, K. S. Madhukar Reddy, S. D. Flexner, C. J. Palmstrøm and P. A. Crowell, "Electrical detection of spin transport in lateral ferromagnet–semiconductor devices," *Nature Phys.,* vol. 3, pp. 197-202, 2007.

[23] N. Tombros, C Jozsa, M. Popinciuc, H. T. Jonkman and B. van Wees, " Electronic spin transport and spin precession in single graphene layers at room temperature," *Nature,* vol. 448, pp. 571-575, 2007.

[24] H. Dery and L. J. Sham, "Spin Extraction Theory and Its Relevance to Spintronics," *Phys. Rev. Lett.,* vol. 98, p. 046602, 2007.

[25] A. N. Chantis, K. B. Belaschenko, D. L. Smith, E. Y. Tsimbal, M. van Schilfgaarde and R. C. Albers., "Reversal of Spin Polarization in Fe/GaAs (001) Driven by Resonant Surface States: First-Principle Calculations," *Phys. Rev. Lett.,* vol. 99, p. 196603, 2007.

[26] G. Kioseoglou, A. T. Hanbicki, R. Goswami, O. M.J . van 't Erve, C. H. Li, G. Spanos, P. E. Thompson and B. T. Jonker, " Electrical spin injection into Si: A comparison between Fe/Si Schottky and Fe/Al2O3 tunnel contacts," *Appl. Phys. Lett.,* vol. 94, p 122106, 2009.

[27] S. Takahashi and S. Maekawa, "Spin current, spin accumulation and spin Hall effect," *Sci. Techn. Adv. Mater.,* vol. 9, p. 014105, 2008

[28] B. J. Jönsson-Åkerman, R. Escudero, C. Leighton, S. Kim, I. K. Schuller, and D. A. Rabson, "Reliability of normal-state current–voltage characteristics as an indicator of tunnel-junction barrier quality," *Appl. Phys. Lett.*, vol. 77, pp. 1870-1872, 2000.

[29] W. F. Brinkman, R. C. Dynes, and J. M. Rowell, "Tunneling Conductance of Asymmetrical Barriers," *J. App. Phys.,* vol. 41, pp. 1915-1921, 1970.

[30] G. Schmidt, D. Ferrand, L. W. Molenkamp, A. T. Filip and B. J. van Wees, "Fundamental obstacle for electrical spin injection from a ferromagnetic metal into a diffusive semiconductor," *Phys. Rev. B,* vol. 62, pp. R4790-R4793 2000.

[31] E. I. Rashba, "Theory of electrical spin injection: Tunnel contacts as a solution of the conductivity mismatch problem," *Phys. Rev. B,* vol. 62, pp. R16267-R16270, 2000.

[32] A. V. Pohm, B. A. Everitt, R. S. Beech and J. M. Daughton., "Bias field and end effects on the switching thresholds of "pseudo spin valve" memory cells", *IEEE Trans. on Magn.,* vol. 33, pp. 3280-3282, 1997.

[33] T. Kimura, Y. Otani and P.M. Levy, "Electrical Control of the Direction of Spin Accumulation", *Phys. Rev. Lett.,* vol. 99, p166601, 2007.

[34] B. C. Min, K. Motohashi, C. Lodder and R. Jansen.," Tunable spin-tunnel contacts to silicon using low-work-function ferromagnets," *Nature Mat.*, vol 5, pp. 817-822, 2006.

[35] P. Mavropoulos, "Spin injection from Fe into Si(001): Ab initio calculations and role of the Si complex band structure," *Phys. Rev. B,* vol. 78, p. 054446, 2008.